# Traffic Flows Analysis in High-Speed Computer Networks Using Time Series

Ginno Millan
Facultad de Ingeniería y Tecnología, Universidad San Sebastian, Lago Panguipulli 1390, Puerto Montt, Chile
ginno.millan@uss.cl

# Traffic Flows Analysis in High-Speed Computer Networks Using Time Series

**ABSTRACT**

This article explores the required amount of time series points from a high-speed traffic network to accurately estimate the Hurst exponent. The methodology consists in designing an experiment using estimators that are applied to time series, followed by addressing the minimum amount of points required to obtain accurate estimates of the Hurst exponent in real-time. The methodology addresses the exhaustive analysis of the Hurst exponent considering bias behavior, standard deviation, mean square error, and convergence using fractional gaussian noise signals with stationary increases.

Our results show that the Whittle estimator successfully estimates the Hurst exponent in series with few points. Based on the results obtained, a minimum length for the time series is empirically proposed. Finally, to validate the results, the methodology is applied to real traffic captures in a high-speed network based on the IEEE 802.3ab standard.

## 1. Introduction

Fractal processes are indicative of stochastic behavior that is invariant to changes in dimensional scales and temporary [1]–[5]. These processes are applied as models in various fields of science [6]–[10]. In the area of telecommunications systems and computer networks, they are used to model traffic on LAN, MAN, WAN, WWW, and different technologies of cellular and wireless networks [11]–[15].

In all these studies, traffic is measured and then analyzed to determine whether or not it fits a fractal behavior. The traffic traces used in all analyzes are made up of hundreds of thousands of samples and often with long capture times. For offline traffic studies [16] these lengths and waiting times are acceptable, however for real-time network administration applications with quality of service metrics based on the precise estimation of Hurst exponent *(H)* [17] these lengths and capture times are too large.

This article studies the behavior of the estimators applied to short-term time series and then addresses the problem of the minimum length required to obtain accurate estimates of *H* in real-time. Therefore, the aim is to obtain high precision with a minimum length, as opposed to [18], where the author raises the impossibility of determining a minimum length for fractal time series without losing their intrinsic properties.

To develop the problem posed by the estimation of *H*, the accuracy must be comparable with the problems of long series; where the accuracy is based on metrics such as bias (*b*), standard deviation (σ), and the mean squared error (*MSE*). Convergence analysis is considered a useful tool to achieve the above.

The article then addresses the following problem: Given the specification(s) of *b* or *MSE*, what should be the minimum length $N_{min}$, of the time series that satisfies them?; that is, suppose that the stochastic process *X* has a *H*, the minimum length $N_{min}$ of the time series must be found so that each proposed $N_{min}$ presents the estimated *H* that is similar to the *H* of the original process *X*.

## 2. Theoretical Framework

*2.1. Fractal Processes.* A process is fractal if its distribution of probabilities is invariant to temporal dilation and compression of its amplitude.

Let $Y = \{Y_t\}_{t \in I}$, where $I = \mathbb{R}$ or $\mathbb{R}^+$, is a real-value stochastic process. It is said that *Y* is a fractal process if and only if there is $H \in \mathbb{R}$, such that for all $a \in \mathbb{R}^+$, the relationship $\{Y_{at}\}_{t \in I} =_d \{a^H Y_t\}_{t \in I}$ is fulfilled, where $=_d$ means equality in their probability distributions [19]. Generally, interest is focused on fractal processes with stationary type increasing with $H > 0$. This definition is known as strict.

A second definition is obtained by invariance in second-order statistics. Let $Y_t$ be a continuous-time stochastic process. $Y_t$ is said to be a second-order self-similar process, if and only if it complies with $E(Y_t) = a^{-H} E(Y_{at})$, for all $a > 0$, $t \geq 0$, and $0 < H < 1$, where $E(\cdot)$ is the median of the process, $\text{Var}[Y_t] = a^{-2H} \text{Var}[Y_{at}]$, for all $a > 0$, $t \geq 0$, and $0 < H < 1$, where $\text{Var}[\cdot]$ is the variance of stochastic process *Y*, and its autocorrelation function, $R(\cdot)$, behaves according to the relationship $R_y(t, s) = a^{-2H} R_y(at, as)$ for all $a > 0$, $t \geq 0$, and $0 < H < 1$ [19].

Computer networks require a discrete version of the definition of fractal processes. The model is defined as $X = \{X_t\}_{t \in \mathbb{Z}}$ which a discrete process due to the sampling of a continuous random signal. *X* is strictly self-similar, if and only if, $X =_d$

$m^{1-H}\Gamma_m(X)$, with $0 < H < 1$, for all $m \in \mathbb{N}$, where $\Gamma(\cdot)$ represents the block aggregation process that receives a time series of length $N$ as input and provides an output as a time series of length $N/m$ [20].

The second version of this definition is that of second-order self-similarity in the exact sense. Formally, $X$ is an exact second-order self-similar process with $\sigma^2$ variance (dispersion) of the process, if its autocovariance function $\rho(k)$ has the following form for the range $0.5 < H < 1$ [21].

$$\rho(k) = \frac{1}{2}\sigma^2[(k+1)^{2H} - 2k^{2H} + (k-1)^{2H}], \text{ for all } k \geq 1 \quad (1)$$

A stochastic process with an autocovariance function given by (1) also satisfies the following restrictions.

$$\text{Var}[X] = m^{2-2H}\text{Var}[\Gamma(X)] \quad (2)$$

$$\text{Cov}[\Gamma_m(X_t), \Gamma_m(X_{t+k})] = m^{2-2H}\text{Cov}[X_t, X_{t+k}] \quad (3)$$

For its variance and covariance, respectively.

In the field of computer networks, a relaxed version of (1) is used. A stochastic process $X$ is asymptotically self-similar to the second order if the correlation factor $\Gamma_m(X)$ when $m \to \alpha$ is equal to the autosimilar stochastic process of discrete-time, that is to say (1).

In particular, let $r(k) = \rho(k)/\alpha^2$ denote the autocorrelation function. For $0 < H < 1$, $H \neq 0.5$, it holds.

$$r(k)H(2H-1)k^{2H-2}, k \to \alpha \quad (4)$$

In particular, if $0.5 < H < 1$, $r(k)$ asymptotically behaves as $ck^{-\beta}$ for $0 < \beta < 1$, where $c > 0$ is a constant, $\beta = 2-2H$, and we have [21].

$$\sum_{k=-\alpha}^{\alpha} r(k) = \alpha \quad (6)$$

That is, the autocorrelation function decays slowly (hyperbolically) which is the essential property that causes it to not be summable. When $r(k)$ decays hyperbolically so that condition (5) holds, we call the corresponding stationary process $X$ LRD [21].

2.2. *On the Hurst Exponent Estimation.* Different methods have been proposed to estimate $H$; these can be classified into methods developed for time domain, frequency domain, and time-scale methods.

Among the time domain methods is the R/S statistic, the aggregated variance method, absolute value method, variance of the residuals, the Higuchi´s method, the scale window variation, the modified variance of allan (MAVAR), Whittle estimator, etc. [22].

The Geweke and Porter-Hudak method, the Periodogram method and the modified Periodogram method are in the frequency domain class which takes advantage of the characteristic power-law behavior of the self-similar processes in the neighborhood of their origin [22].

Finally, time-scale methods include wavelet-based estimators such as the Abry and Veitch´s method [22], [23].

## 3. Working Methodology

The study of fractality by analyzing the value of H allows its presence and its degree of persistence to be detected. The methodology developed for the calculation of the different estimators for different lengths of time series is described by the follow Figure 1.

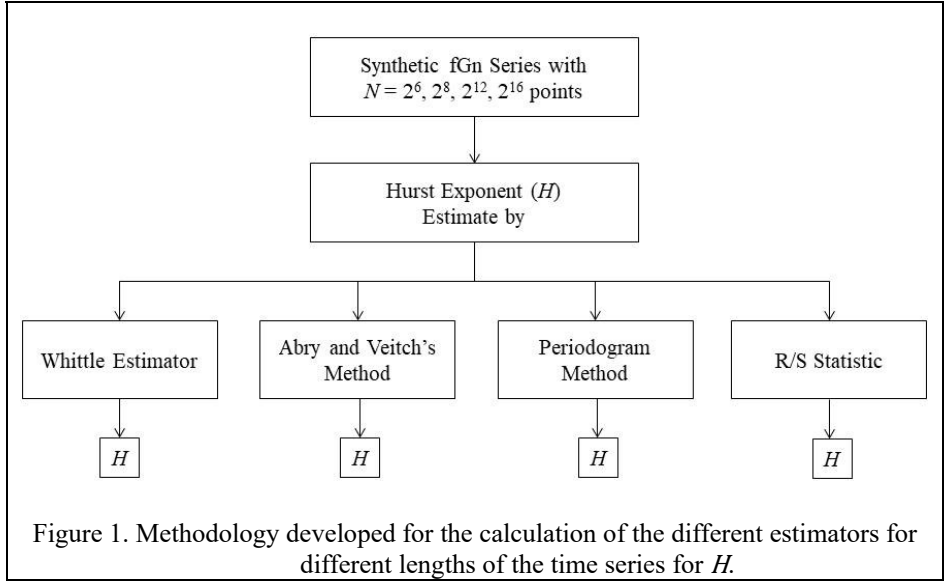

Figure 1. Methodology developed for the calculation of the different estimators for different lengths of the time series for H.

To apply the estimators, the $N$ time series must be obtained. Synthetic signals with known $H$ are obtained through the simulation of a series of fractional Gaussian noise (fGn) [24] with stationary increases using the Davies and Harte method described in detail in [25].

For the experiments, 200 traffic traces with $H \in \{0.5, 0.6, 0.7, 0.8, 0.9\}$ and lengths $N = \{2^i, i = 6, 7, 8, 9, 10, 11, 12, 13, 14, 15, 16\}$ were, considered, that is 11000 fractal signals (100 * $H$ * $N$). For each of the sets of estimates of a particular $H$, the calculation of the statistics described above was performed:

→ Bias $b = H_0 - \bar{X}$ where $H_0$ is the nominal value of $H$ and $\bar{X}$ is the average of the values of $X$ process.

→ Standard deviation σ.

→ $MSE = N^{-1} \sum_{i=1}^{N} (x_{i-H_0})^2$

Then, based on these three estimators, a minimum series length, $N_{min}$, is proposed from the estimates that consider $b \sim 0.03$ and $\sigma \sim 0.015$.

Together with the above, estimates based on b and σ are classified as follows.

→ High precision: when $b \leq 0.03$ and $\sigma \leq 0.015$.
→ Acceptable: when $0.03 < b < 0.05$ and $\sigma \leq 0.02$.
→ Biased (but not unacceptable): when $b > 0.1$.

Once the minimum length for the fGn series is obtained, the results obtained are then applied to real traffic traces. For these series, designated as $Z$, of length $M$, such that $M \gg N_{min}$ the procedure is:

1. Let $t_0,..., t_k$ be a sequence of points on the x-axis, where it is true that $t_{j+1} > t_i$ and $(t_{i+1}-t_i) < N_{min}$, for each block of $Z$ of length $N_{min}$, $\{Z_j\}_{j=t_i}^{t_i+N_{min}-1}$, be the estimate of $H$, that is, $H_{e,t_i}^{N_{min}}(\cdot)$, until $t_k + N_{min} > M$ for any $k$.
2. It is speculated that $N_{min}$ is chosen correctly, if $t_i$ versus $H_{e,t_i}^{N_{min}}(\cdot)$ is plotted, the result show a signal with a little variation, that is, the variation should be equal to the value of $\sigma$. This is explained as follows: The correct length $N_{min}$ is directly related to the convergence of a series, for this reason, the study of this relationship is carried out as follows, but not before remembering that the convergence of an estimator is obtained by disaggregating the original $Z$ series in blocks of size $m \ll M$, to obtain a set $Z = \{\Psi_1^m, \Psi_2^m, ..., \Psi_i^m\}$, where the set $\Psi_i^m$ is defined by $\Psi_i^{m,} = \{Z_{(i-1)}, Z_{(I-m)m+1}, ..., Z_{im}\}$.
3. $H_{e,t_i}^{N_{min}}(\cdot)$ is applied over a joint set of the series, i.e. $\bigcup_{j=1}^{Nm^{-1}}\{\Psi_i^m\}$, to obtain a sufficient amount of $H$ estimators; amount described by $H_{e,1}^m, H_{e,1}^{2m},..., H_{e,1}^{jm}$.
4. Finally, $H$ versus $j_m$ is plotted, for $j = 1, 2,..., Mm^{-1}$, to visually verify the convergent behavior of the estimator $H_{e,t_i}^{N_{min}}(\cdot)$ if the speculation is correct.

## 4. Simulation and Results

*4.1. Analysis of the estimators:* The main results are presented below.

Figure 2 presents the estimate of $H$ for a series of synthetic fGn with length $N=2^6$ points (a), $N=2^8$ points (b), $N=2^{12}$ points (c), and $N=2^{16}$ points *(d)* using the Whittle estimator incorporated in SELFIS [26], [27].

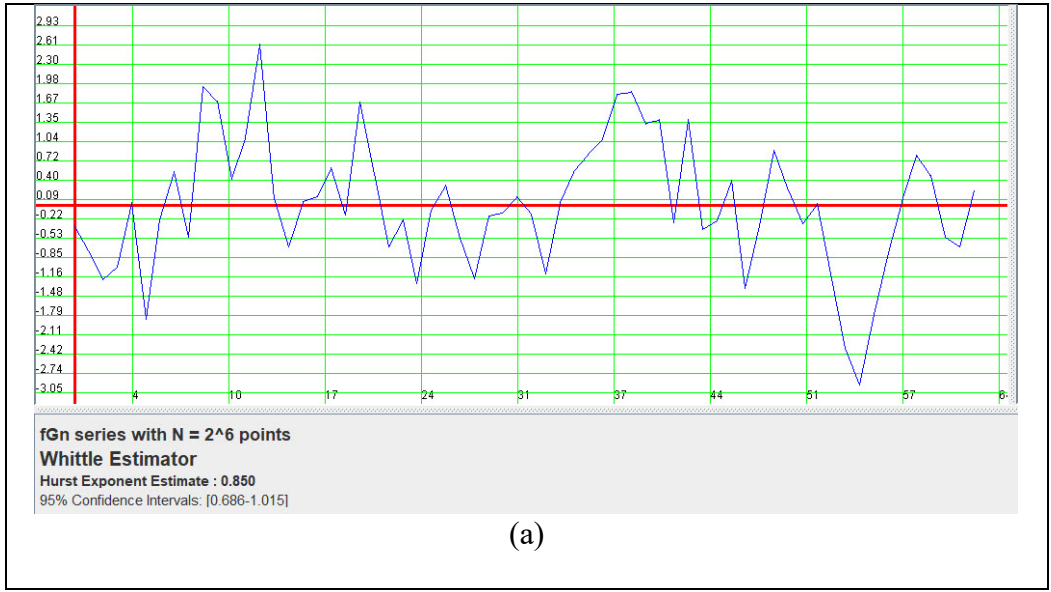

(a)

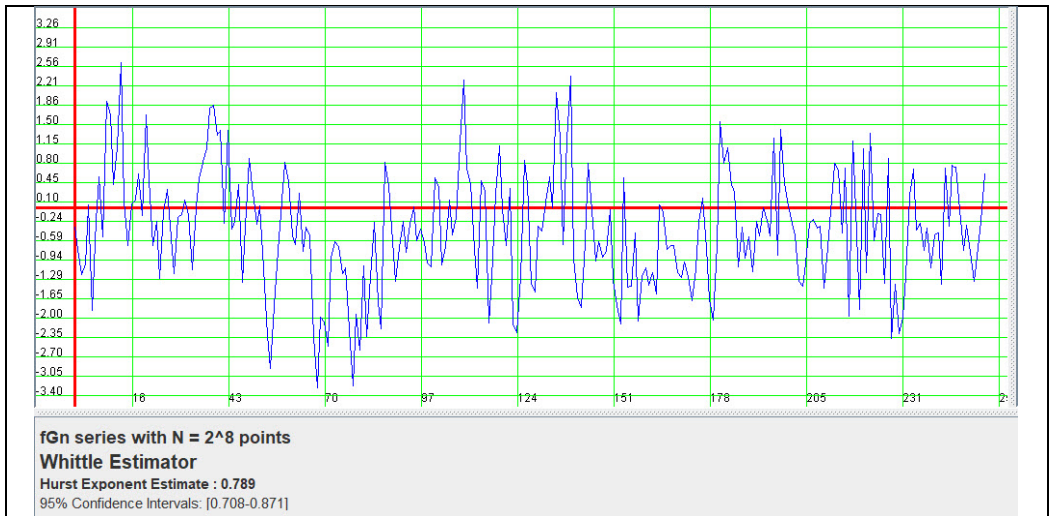

(b)

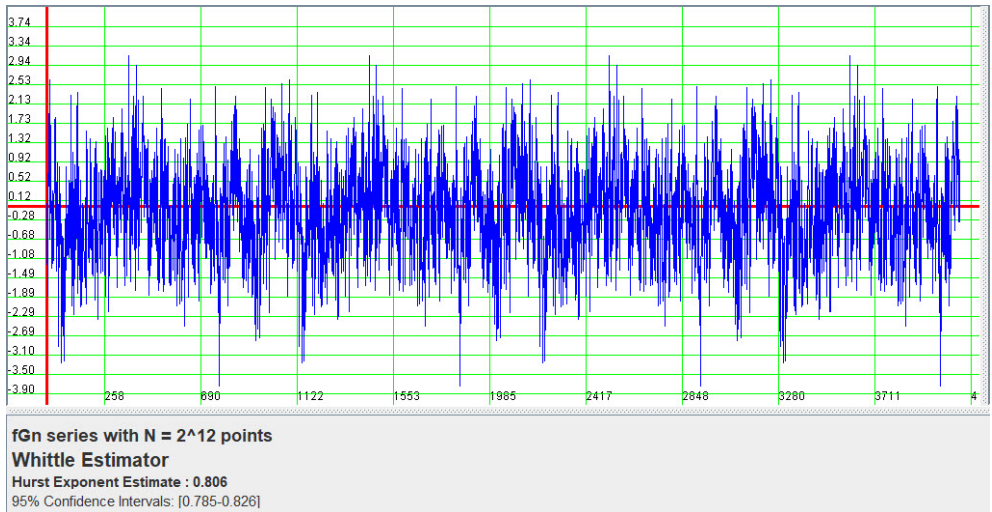

(c)

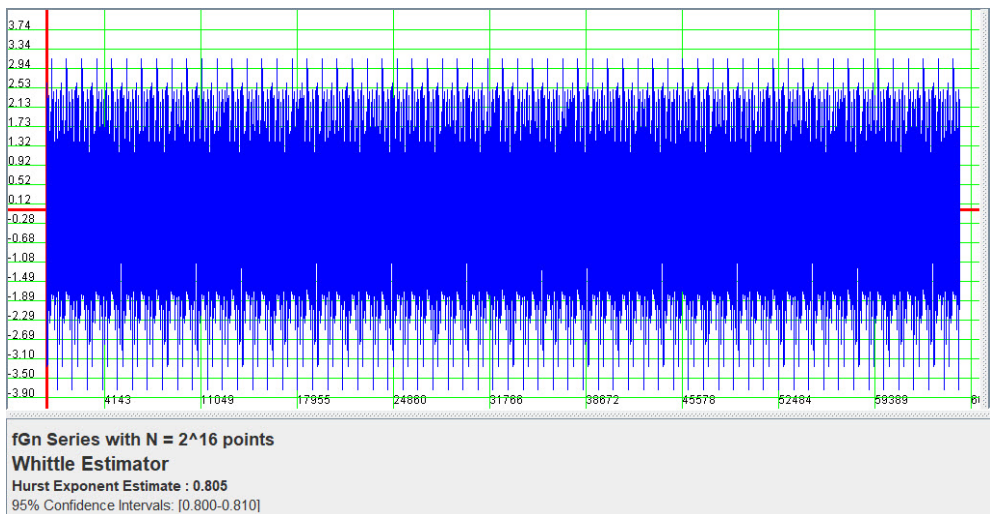

(d)

Figure 2. (a) Synthetic fGn series with $N = 2^6$, (b) $N = 2^8$, (c) $N = 2^{12}$, and (d) $N = 2^{16}$ points, respectively, for which the value of $H$ is estimated using the Whittle estimator.

Figure 3 presents the estimate of *H* for the same fGn series using the Abry and Veitch´s method that incorporates the same application.

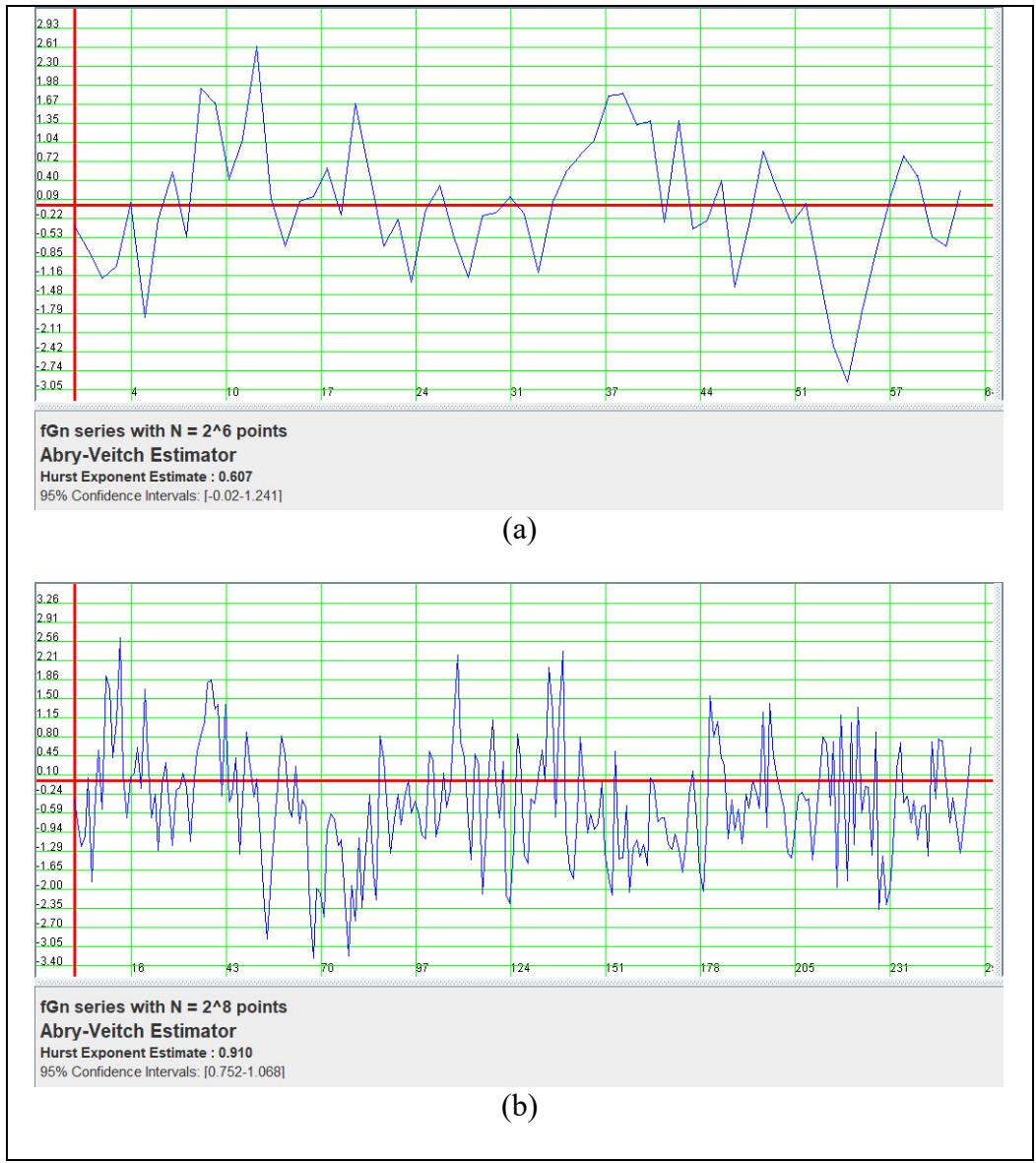

(a)

(b)

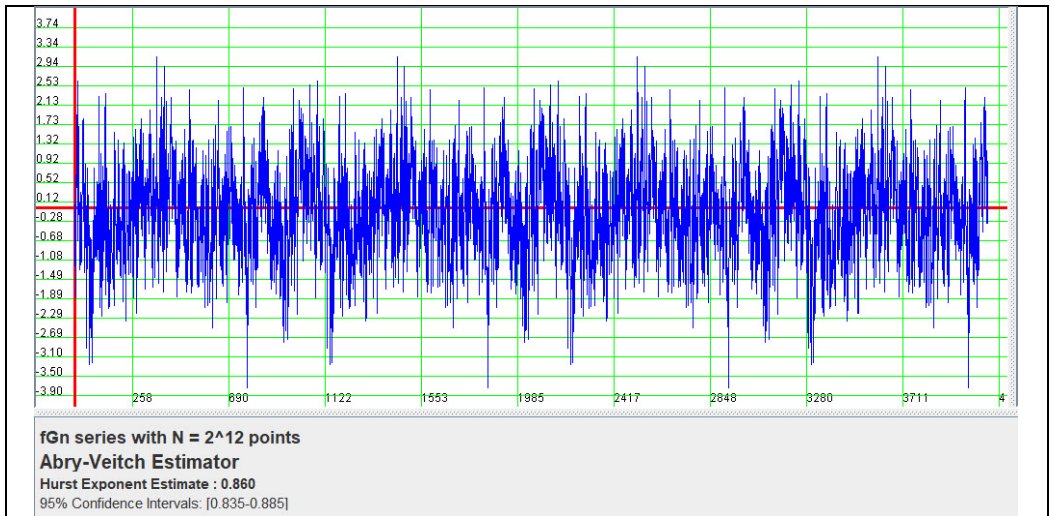

(c)

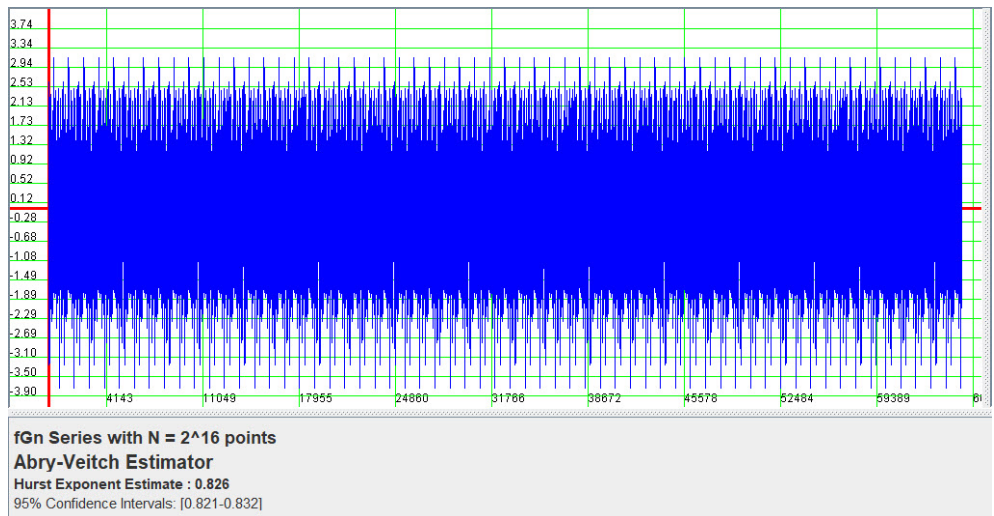

(d)

Figure 3. (a) Synthetic fGn series with $N= 2^6$, (b) $N= 2^8$, (c) $N= 2^{12}$, and (d) $N= 2^{16}$ points, respectively, for which the value of $H$ is estimated using the Abry and Veitch´s method.

Figure 4 shows the behavior of the previous traces when applying the Periodogram method. The Periodogram method finds its basis in the behavior of the near origin of the power spectral density (PSD).

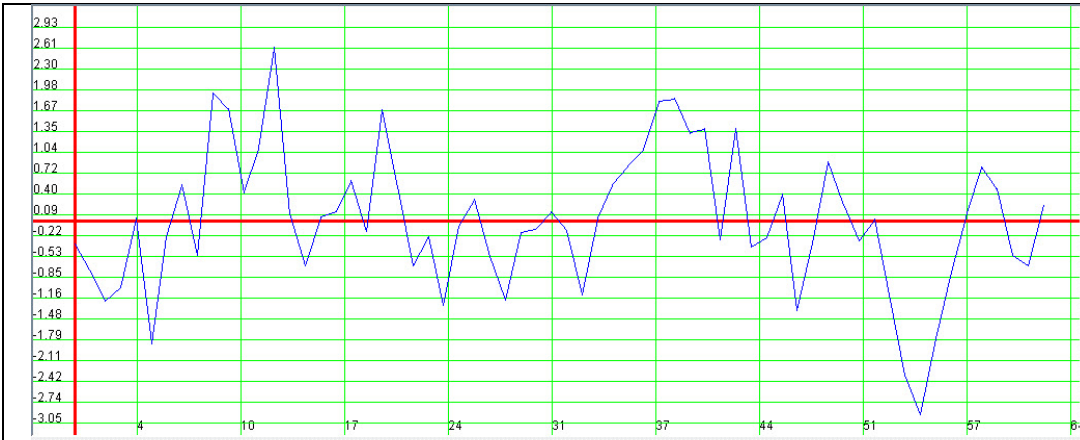
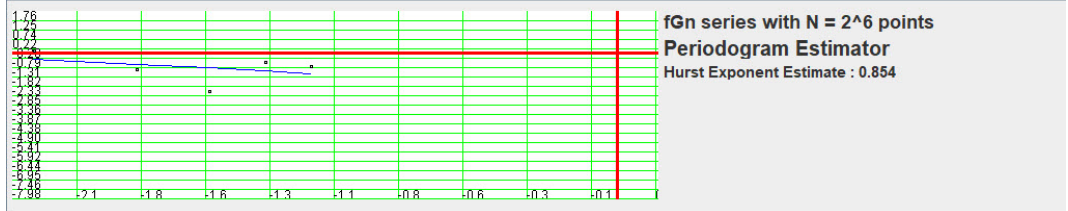

(a)

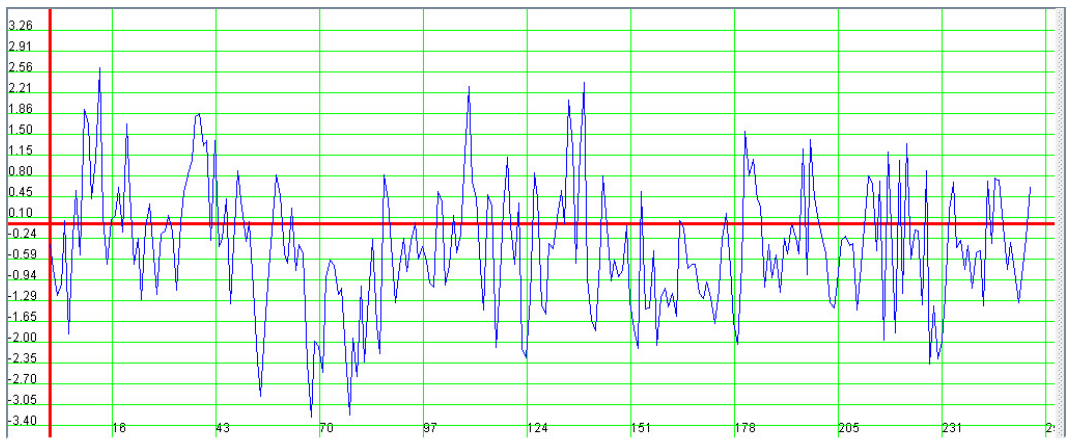
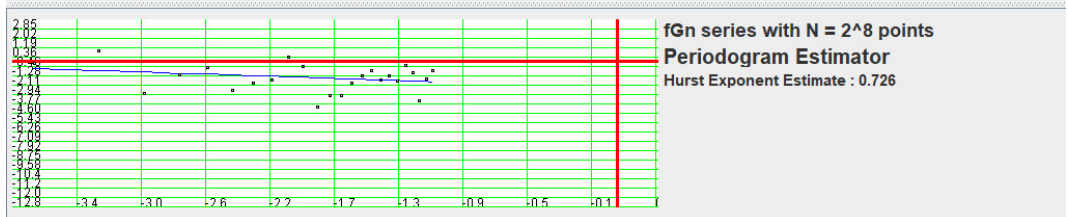

(b)

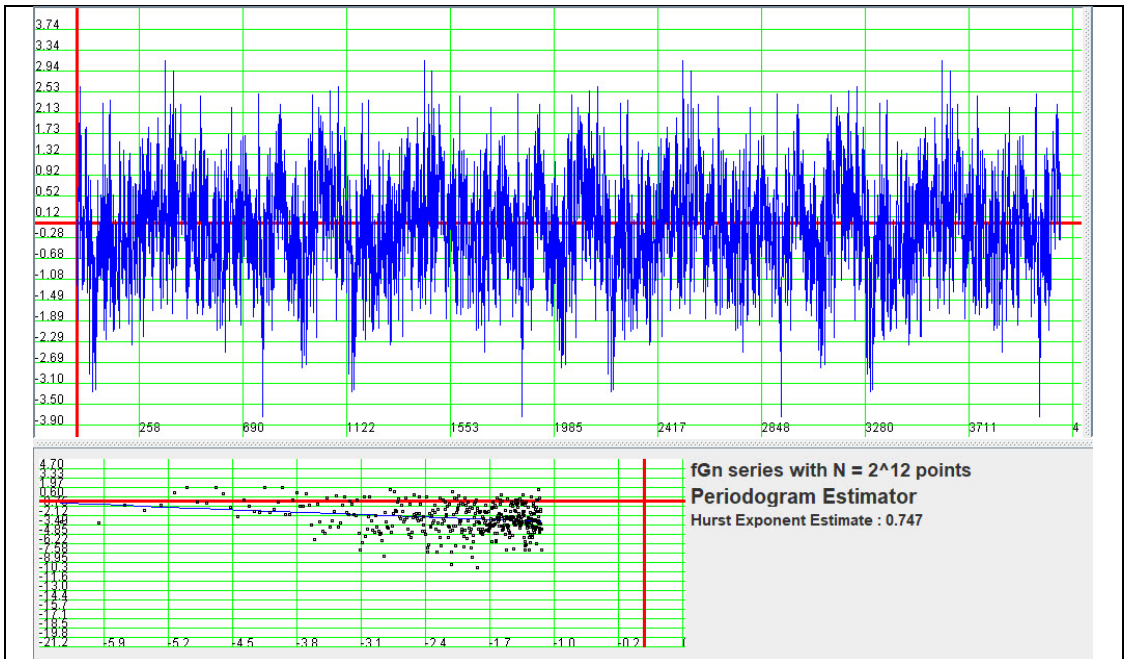

(c)

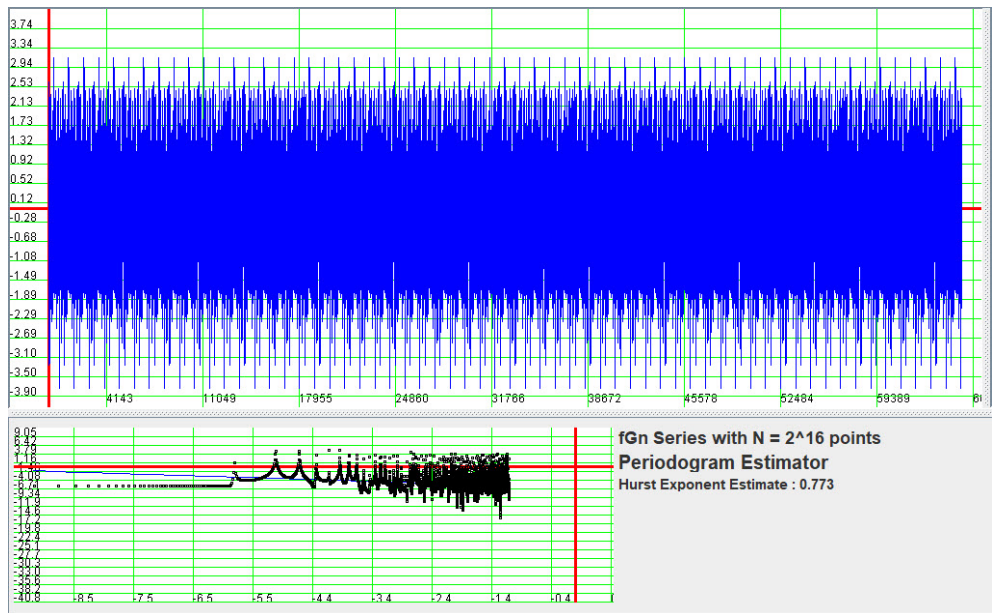

(d)

Figure 4. (a) Synthetic fGn series with $N = 2^6$, (b) $N = 2^8$, (c) $N = 2^{12}$, and (d) $N = 2^{16}$ points, respectively, for which the value of $H$ is estimated using the Periodogram method.

Figure 5 shows the behavior of the previous traces when applying the R/S statistic. This graphic method has high variability regardless of the length of the series.

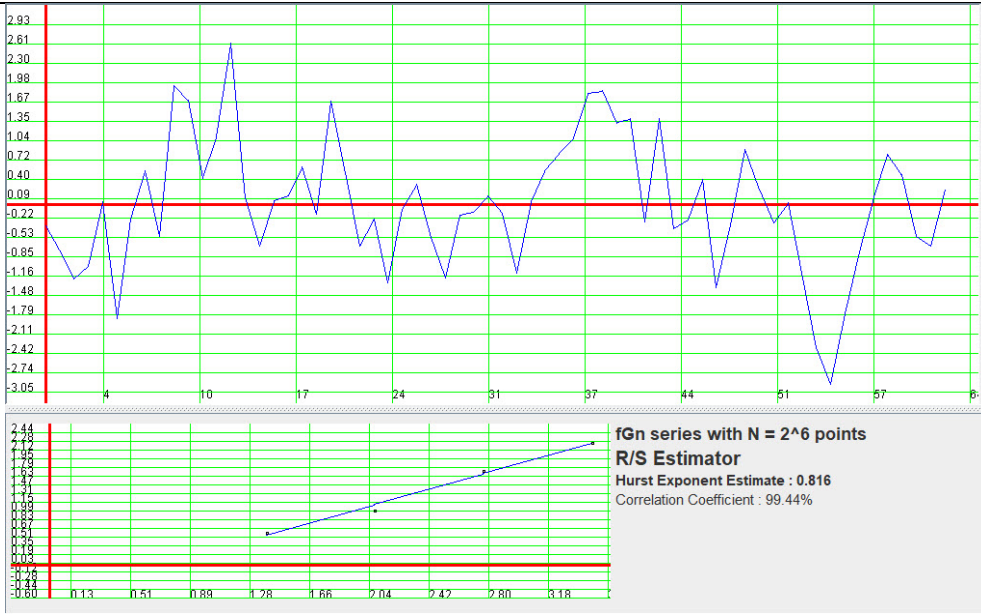

(a)

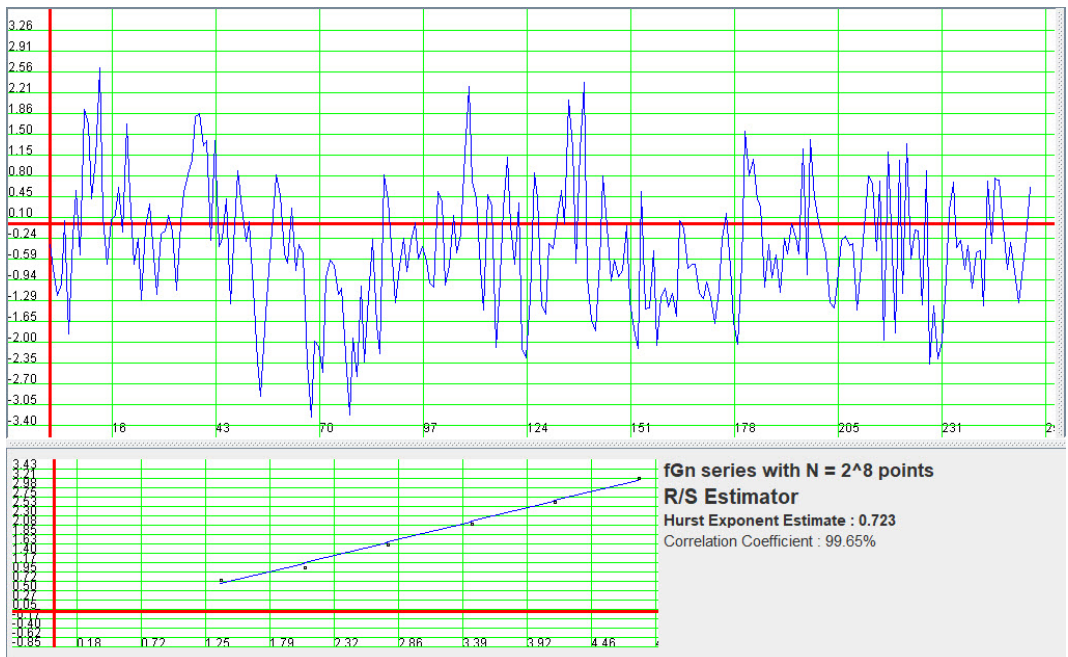

(b)

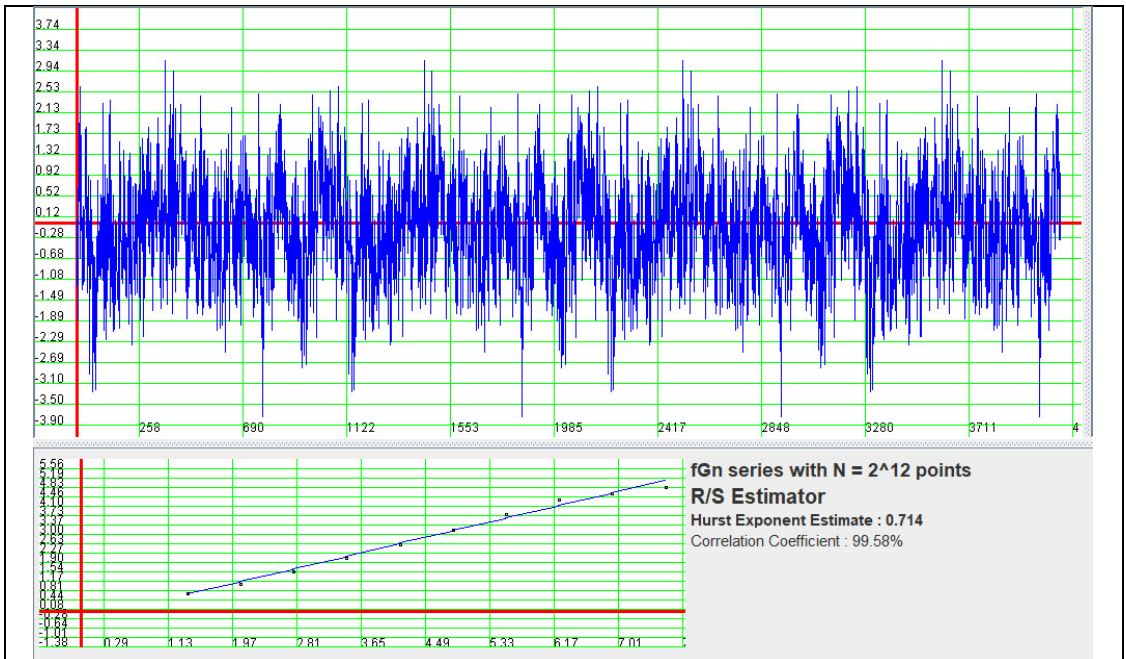

(c)

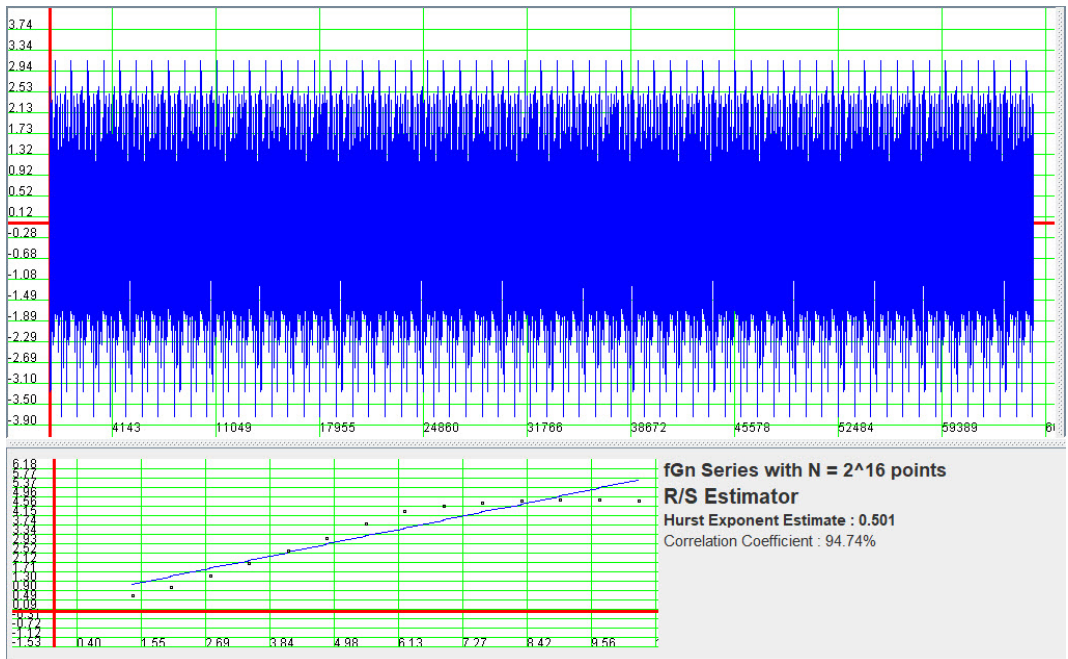

(d)

Figure 5. (a) Synthetic fGn series with $N = 2^6$, (b) $N = 2^8$, (c) $N = 2^{12}$, and (d) $N = 2^{16}$ points, respectively, for which the value of $H$ is estimated using the R/S statistic.

Table 1 shows a summary of the obtained results.

| Used Technique | H value for the synthetic fGn series with | | | | Respective Figure |
|---|---|---|---|---|---|
| | $N=2^6$ | $N=2^8$ | $N=2^{12}$ | $N=2^{16}$ | |
| Whittle estimator | 0.850 | 0.789 | 0.806 | 0.805 | 2(a), 2(b), 2(c), 2(d) |
| Abry and Veitch´s method | 0.607 | 0.910 | 0.860 | 0.826 | 3(a), 3(b), 3(c), 3(d) |
| Periodogram method | 0.854 | 0.726 | 0.747 | 0.773 | 4(a), 4(b), 4(c), 4(d) |
| R/S statistic | 0.816 | 0.723 | 0.714 | 0.501 | 5(a), 5(b), 5(c), 5(d) |

Table 1. Summary of the results obtained in Figures 2, 3, 4, and 5.

Based on the results obtained from the experiments for $2^6 < N < 2^{16}$, the estimates of $H$ deliver acceptable results, i.e. $0.03 < b < 0.05$ and $\sigma \leq 0.02$ and for $N \geq 2^{13}$ the estimates are high precise, i. e. $r \sim 0.03$ and $\sigma \sim 0.015$.

The analysis of the results obtained shows that the estimates of $H$ exponent using the Whittle estimator show high precision and low variability for the standard length series reported in the literature, for $N \sim 2^{10}$ points. Based on the analysis of the results obtained if $N < 2^{12}$ and $H \geq 0.8$, the estimates are accurate.

On the other hand, the results obtained using the Whittle´s estimator are compared with the results obtained using the Abry and Veitch´s method. The results based on Whittle´s estimator are more accurate than their wavelet counterparts for short series in the context of the synthetic fGn series of the study.

Figure 6 shows the bias for all the methods used on the fGn series for different data lengths. This allows us to observe the variation of the different estimates of value $H$ delivered by each analysis method.

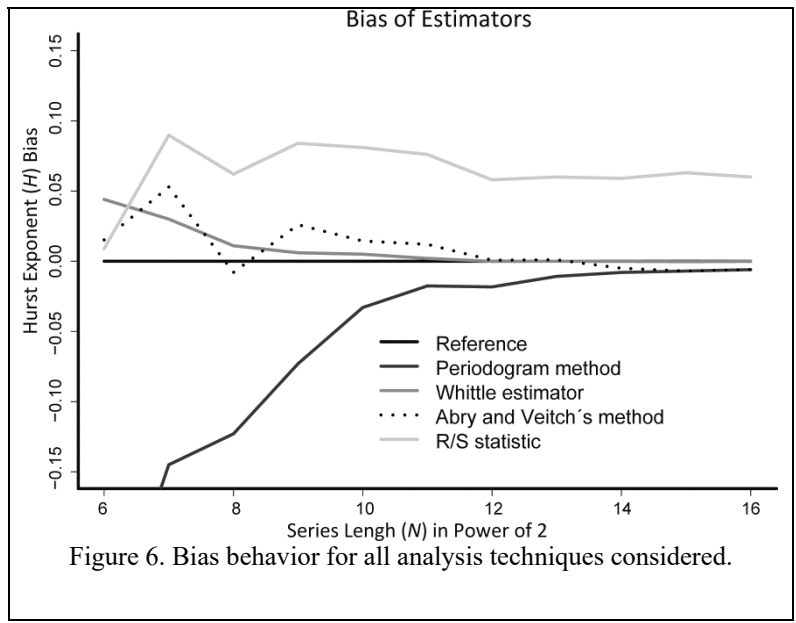

Figure 6. Bias behavior for all analysis techniques considered.

Note from in Figure 6 that the Whittle´s estimator and Abry and Veitch´s method have a better behavior for short length series, $N < 2^{10}$ points, than the other analysis techniques. Also, the Whittle estimator behaves with less irregularity than Abry and Veitch´s method for short length series ($N < 2^{10}$) and for series with lengths $N \geq 2^{10}$, the bias presented by both techniques is not significant.

For the Abry and Veitch´s method, the bias behaves irregularly for the short series and stabilizes with $N \geq 2^{14}$. For the Whittle estimator, the bias behaves irregularly for the short series and stabilizes with $N \geq 2^{11}$. The other techniques (Periodogram Method and R/S statistic) exhibit a completely irregular behavior and a very high bias and unless $N \geq 2^{16}$ (see

Figure 6), which is why its estimates are not considered acceptable. The bias of the R/S statistic does not show behavior that stabilizes, while the Periodogram method is stabilized for a high $N$, that is, $N \geq 2^{16}$ (see Figure 6).

With the same objective pursued by the study of the bias of the estimators considered in Figure 6, Figure 7 illustrates the behavior of the standard deviation of the estimators for the traces with variable $N$ length in the interval $2^6 < N < 2^{16}$.

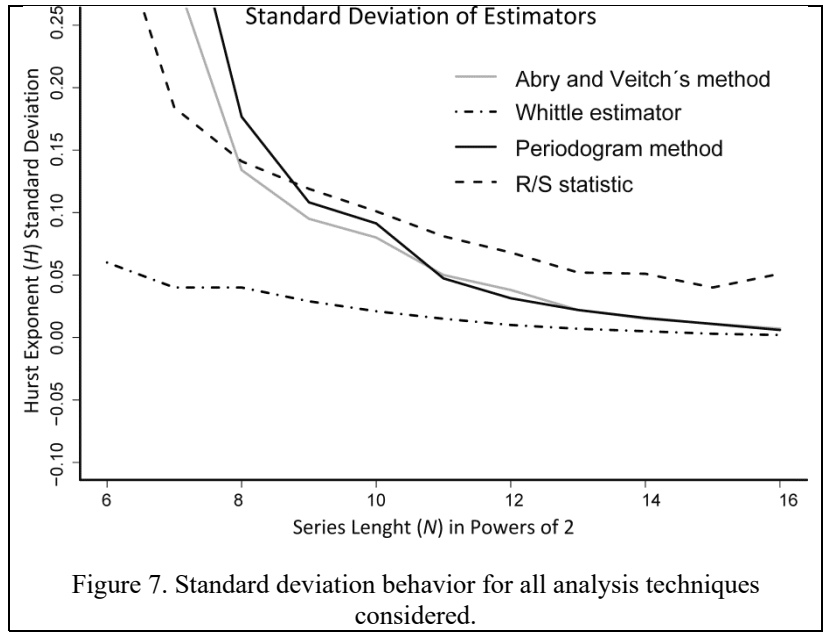

Figure 7. Standard deviation behavior for all analysis techniques considered.

Note from Figure 7 that the Whittle estimator is the one that presents the least variability and for $N \geq 2^8$. the estimates are not precise enough to be considered. Abry and Veitch´s method and the Periodogram method follow in precision, and the length required for the series can be considered as relatively identical when considering variability.

Finally, when considering both bias (Figure 6) and standard deviation (Figure 7), the best estimator for short time series is the Whittle estimator, with high precision results for $N \geq 2^8$, the Abry and Veitch´s method presents good precision for N ~ $2^{12}$, the Periodogram method has acceptable estimates for N > $2^{15}$ and the R/S statistic has very biased estimates for $2^6 < N < 2^{16}$.

*4.2. Mean Convergence Analysis of the Estimators:* In the methodology section specifies how to obtain $N_{min}$ length for the fGn series. The results obtained are applied to real traffic traces. Traffic traces are designated by $Z$ and have a length $M$ such that M $\gg N_{min}$. The analysis presented establishes the need to analyze the convergence of the estimators by the following series of steps:

1. A length $M = 2^{16}$ is arbitrarily chosen for each series to study the convergence of each of them.
2. The convergence analysis is initially applied to the first $t_0 = 2^6$ points of the series $X_j$, that is, the estimator of $H$, $H_{e,t_i}^{N_{min}}(\cdot)$, is applied to the first $t_0$ points of $X_j$.
3. The estimator is applied repeatedly to the following points of the series, that is to say the following $t_0 + it_u$ points of $X_j$, where $t_u = 200$ and i = 1, 2,..., k, such that $t_0 + kt_u \leq M$.
4. The analysis is applied to 200 series of fGn to obtain a graph of the convergent behavior of each of the series.

5. Once the convergence analysis has been developed for each series of the previous point, an average convergence analysis is performed. This means obtaining the average for the 200 estimates of $t_0$ and so on until all points of each fGn series are completed.

With the previous steps the average convergence graph resulting from $H_{e,t_i}^{N_{min}}(\cdot)$ versus $t_i$, $i = 0, 1,..., k$ now results in an indicator of how well the techniques considered convergence to the theoretical value of $H$. Therefore, this graph showing the average convergent value is applied to all the estimation techniques of $H$ considered in this investigation.

Figure 8 shows the mean convergent behavior of R/S statistic. The R/S statistic quickly stabilizes, however it is biased by 0.05; this level of uncertainty does not allow a minimum length of the time series to be defined.

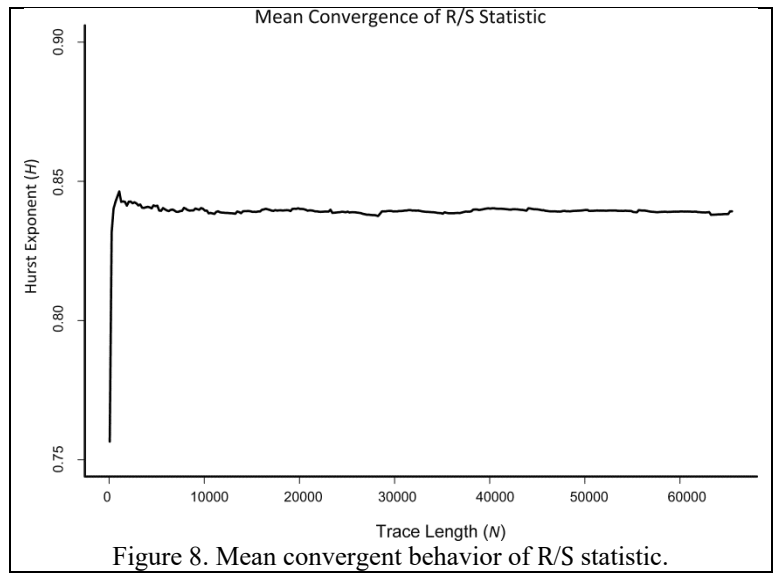

Figure 8. Mean convergent behavior of R/S statistic.

Figure 9 shows the mean convergent behavior of the Whittle estimator, with rapid stabilization and low bias. From this perspective, its behavior is similar to the behavior of the R/S statistic, but with the difference that the estimates of $H$ from the Whittle estimator prove to have a low biased. For precise estimates of $H$, at least a quantity $N \geq 2^9$ points is required.

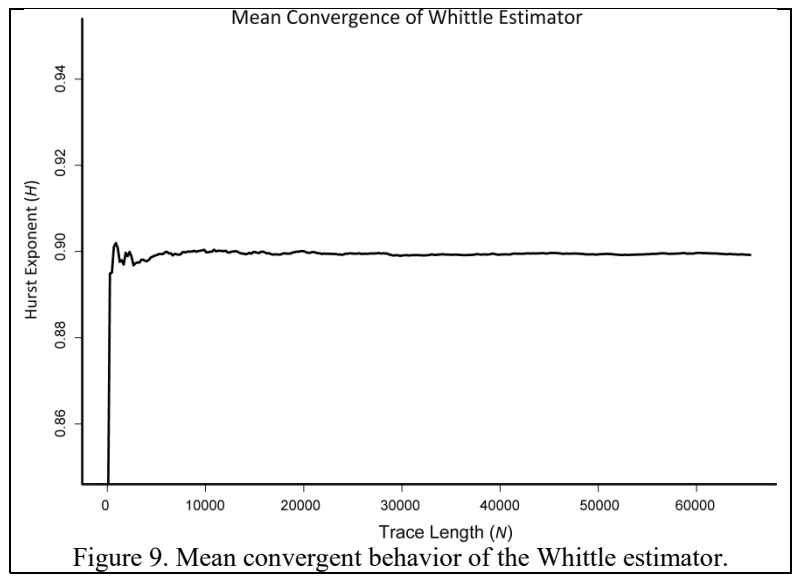

Figure 9. Mean convergent behavior of the Whittle estimator.

Figure 10 shows the mean convergent behavior of the Periodogram method. This method requires $N \geq 2^{13}$ points to achieve accurate estimates.

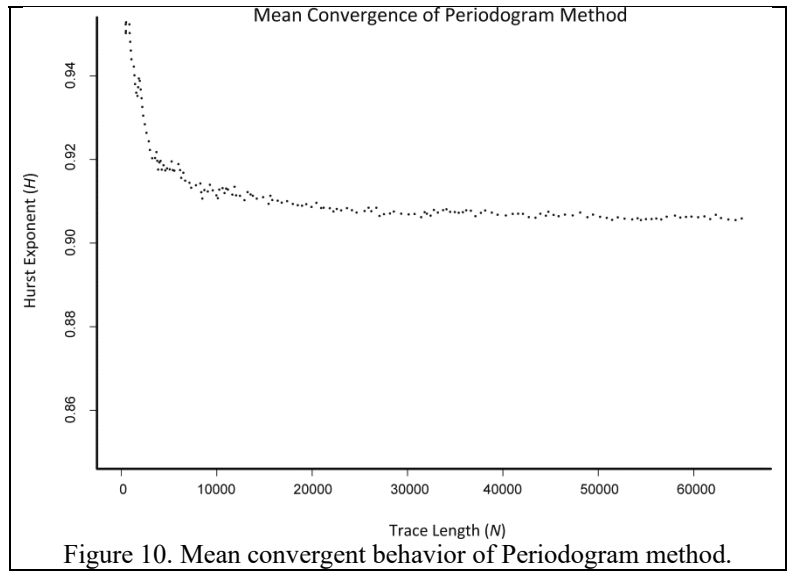

Figure 10. Mean convergent behavior of Periodogram method.

Figure 11 shows the mean convergent behavior of the Abry and Veitch´s method. This method requires $N \geq 2^{12}$ points to achieve accurate estimates.

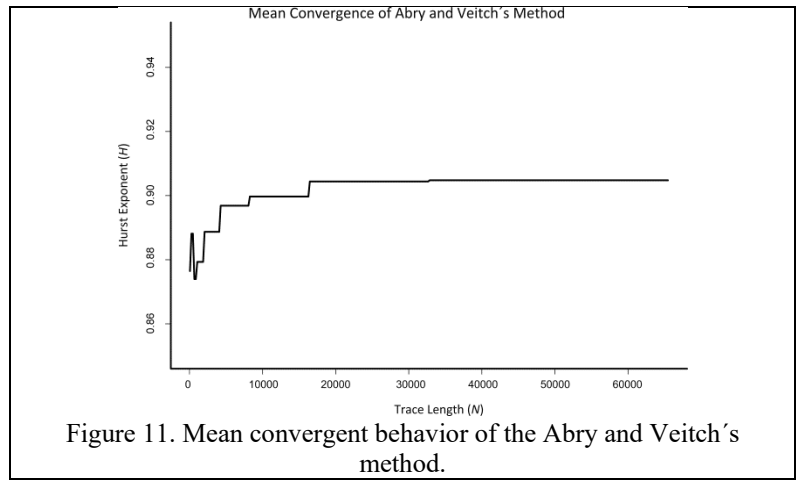

Figure 11. Mean convergent behavior of the Abry and Veitch´s method.

The larger size of the time series for the Abry and Veitch´s method and Periodogram method is explained by the information extracted from Figure 6; the Periodogram method is negatively biased and the Abry and Veitch´s method is in principle positively biased and in the long term it becomes negative.

## 5. Application to Real Traffic Traces

5.1. *Generalities and Specification of the Test Scenario:* Based on the previous results on synthetic fGn traces, partial conclusions were applied to real traffic traces.

The traces were obtained from the core switch from the Electrical Engineering Department of the University of Santiago de Chile with the center of operations control (NOC) of the University corporate network. For the capture of the traces,

Wireshark [28] is used and they are transformed into time series using MatLab, to construct ordered pairs (time of arrival at the sniffer, size of the captured frame). Capture considers, the storage and processing capacity of 3 LAN traffic hours of the Electrical Engineering Department of the University of Santiago de Chile considering bidirectional traffic, which translates into $N = 2^{32}$ points through the following procedure:

1. Let $t_0, t_1, \ldots, t_k$ be a sequence of points on the x-axis, where $t_{i+1} > t_i$ and $(t_{i+1} - t_i) < N_{min}$ is verified. Unless this time $N_{min} = 2^8$ points is defined, which implies starting with a time series given by $t_{i+1} > t_i$ and $(t_{i+1} - t_i) = 2^8$.
2. Each aggregate block in the series it is estimated in reverse, the blocks change to a shorter length to approximate total real traffic.

Considerations regarding points 1 and 2 above:

- The value $N_{min} = 2^8$ results from a general appreciation of the results obtained in the theoretical sections; therefore, it is not a value chosen for convenience of analysis.
- Applying a disaggregation process can suggest, or be interpreted, as a process contradictory to the theory, but this research seeks a $N_{min}$ that for both theoretical and practical results leads to a useful tool in real-time. However, an aggregation procedure cannot be homologated to a real-time one for the estimation of $H$ and the decision making for network administration and their implications and consequences.

5.2. *Obtained Results:* Table 2 shows the general results for the H estimation techniques considered.

Table 2. General overview of estimates of $H$ for real traffic traces.

| Used Technique | Estimated Value of $H$ | Correlation Coefficient | Confidence Intervals |
|---|---|---|---|
| Whittle estimator | 0.806 | – | 95% [0.785 – 0.826] |
| Abry and Veitch method | 0.860 | – | 95% [0.835 – 0.885] |
| Periodogram method | 0.947 | – | – |
| R/S statistic | 0.714 | 99.58% | – |

From Table 2, it is observed that both the Whittle estimator and that Abry and Veitch technique are consistent with the estimates of $H$ reported in [22]. However, for the Abry and Veitch method a high non-accused variability is observed in said reference; while the Periodogram method overestimates the value of $H$ and the R/S statistic exhibits irregular behavior.

Finally, the main results obtained can be grouped as follows:

The Whittle's estimator behaves in a good way when it comes to short series and long series that exhibit both minimal bias and variability.

The Abry and Veitch´s method behaves acceptably when applied to medium length time series.

The Periodogram method behaves acceptably if it is applied to time series of medium length.

The R/S statistic shows a high bias and therefore it is not suitable for application to short-term time series.

The minimum length for the Whittle estimator is $N \geq 2^8$ points.

The minimum length for the Abry and Veitch's method is $N \geq 2^{13}$ points.

The minimum length for the Periodogram method is $N \geq 2^{15}$ points.

It is not possible to estimate a minimum series length for the R/S statistic due to its high bias and its high variability when it comes to lengths of $N \geq 2^{16}$ points.

It is not possible to find a length $N_{min}$ for the time series that responds to all the estimation techniques of H considered in this investigation.

**6. Discussion**

Traditional process-based traffic models with short-range dependency do not provide details on the behavior of flows in current high-speed computer networks. Consequently, it is necessary to rethink the study of computer networks charging models that consider fractal entry traffics, since their requirements impose new challenges to network engineering, especially in buffering strategies of active equipment and estimation of yields.

This study presents the behavior of the most used estimators in the literature in synthetic time series using fGn series and then extrapolates the results to real traffic traces obtained from a high-speed network based on the IEEE 802.3ab standard.

Based on the behavioral study of the time series used of both fGn and real IEEE 802.3ab by applying bias analysis, average convergence analysis, and H estimation, an attempt was made to determine a minimum length called $N_{min}$ to establish a broad criterion of fractal analysis.

Research as such does not result in a specified length, since each type of analysis responds, even with the same objective to different scenarios in which it is a question of checking the fractal nature of traffic flows in high-speed computer networks of different technologies.

**7. Conclusions**

In this study, we presented the analysis of traffic flows in high speed computer networks using a minimum quantity of time series points that must contain estimate the *H*. An experiment using estimators applied to time series provides an accurate determination of *H* in real-time. In the exhaustive analysis of *H*, bias behavior, standard deviation, mean square error, and convergence using fractional gaussian noise signals with stationary increases are considered.

The behavior of most estimators used in the literature, in synthetic time series using fGn series followed by extrapolation, results in real traffic traces obtained from a high-speed network based on the IEEE 802.3ab standard. Based on this behavior by applying bias analysis, average convergence analysis, and *H* estimation, an attempt was made to determine a minimum length called $N_{min}$ and to establish a broad criterion of fractal analysis.

The results obtained with the Whittle estimator allowed the Hurst exponent to be obtained in series with a few points. Then, a minimum length for the time series is empirically proposed.

Finally, to validate the results, the methodology was applied to actual traffic captures in a high-speed network based on the IEEE 802.3ab standard. The Whittle's estimator behaves in a good way when it comes to short series and long series that exhibit both minimal bias and variability.

## References

[1] G. Millán and G. Lefranc, "Presentation of an estimator for the hurst parameter for a self-similar process representing the traffic in IEEE 802.3 networks," *Int. J. Comput. Commun. Control*, vol. 4, no. 2, pp. 137–147, 2009.

[2] R. Fontugne et al., "Scaling in internet traffic: a 14 year and 3 day longitudinal study, with multiscale analyses and random projections," *IEEE/ACM Trans. Netw.*, vol. 25, no. 4, pp. 2152–2165, 2017.

[3] S. H. Sørbye and H. Rue, "Fractional Gaussian noise: prior specification and model comparison," *Environmetrics*, vol. 29, no. 5–6, p. e2457, Aug. 2018.

[4] G. Millán, E. San Juan, and M. Vargas, "A simple multifractal model for self-similar traffic flows in high-speed computer networks," *Comput. y Sist.*, vol. 23, no. 4, pp. 1517–1521, 2019.

[5] Z. Tian, "Chaotic characteristic analysis of network traffic time series at different time scales," *Chaos, Solitons and Fractals*, vol. 130, p. 109412, 2020.

[6] A. Jaegler, "Fractal organization: preparing for the factory of the future," *IEEE Eng. Manag. Rev.*, vol. 46, no. 4, pp. 136–142, 2018.

[7] A. Paramasivam et al., "Influence of electrode surface area on the fractal dimensions of electrogastrograms and fractal analysis of normal and abnormal digestion process," in *IEEE International Conference on Recent Trends in Electrical, Control and Communication*, 2019, pp. 245–250.

[8] Z. Chen, Y. Hu, and Y. Zhang, "Effects of compression on remote sensing image classification based on fractal analysis," *IEEE Trans. Geosci. Remote Sens.*, vol. 57, no. 7, pp. 4577–4590, 2019.

[9] R. K. Rout, S. K. S. Hassan, S. Sindhwani, H. M. Pandey, and S. Umer, "Intelligent classification and analysis of essential genes using quantitative methods," *ACM Trans. Multimed. Comput. Commun. Appl.*, vol. 16, no. 1s, pp. 1–21, 2020.

[10] G. Kermarrec, "On estimating the Hurst parameter from least-squares residuals. Case study: correlated terrestrial laser scanner range noise," *Mathematics*, vol. 8, no. 5, p. 674, 2020.

[11] G. Millán, H. Kaschel, and G. Lefranc, "Discussion of the analysis of self-similar teletraffic with long-range dependence (LRD) at the network layer level," *Int. J. Comput. Commun. Control*, vol. 5, no. 5, pp. 799–812, 2010.

[12] G. Millán and G. Lefranc, "Application of the analysis of self-similar teletraffic with long-range dependence (LRD) at the network layer level," *Int. J. Comput. Commun. Control*, vol. 10, no. 1, pp. 62–69, 2015.

[13] X. Ge et al., "Wireless fractal cellular networks," *IEEE Wirel. Commun.*, vol. 23, no. 5, pp. 110–119, 2016.

[14] J. Chen, X. Ge, and Q. Ni, "Coverage and handoff analysis of 5g fractal small cell networks," *IEEE Trans. Wirel. Commun.*, vol. 18, no. 2, pp. 1263–1276, 2019.

[15] S. Gupta, P. Kshirsagar, and B. Mukherjee, "A low-profile multilayer cylindrical segment fractal dielectric resonator antenna: usage for wideband applications," *IEEE Antennas Propag. Mag.*, vol. 61, no. 4, pp. 55–63, 2019.

[16] A. Callado et al., "A survey on internet traffic identification," *IEEE Commun. Surv. Tutorials*, vol. 11, no. 3, pp. 37–52, 2009.

[17] J. Mielniczuk and P. Wojdyłło, "Estimation of Hurst exponent revisited," *Comput. Stat. Data Anal.*, vol. 51, no. 9, pp. 4510–4525, 2007.

[18] R. Ritke, X. Hong, and M. Gerla, "Contradictory relationship between Hurst parameter and queueing performance (extended version)," *Telecommun. Syst.*, vol. 16, no. 1–2, pp. 159–175, 2001.

[19] O. I. Sheluhin, S. M. Smolskiy, and A. V. Osin, "Principal concepts of fractal theory and self-similar processes," in *Self-Similar Processes in Telecommunications*, Chichester, UK: John Wiley & Sons Ltd, 2007, pp. 1–47.

[20] M. E. Sousa-Vieira, A. Suárez-González, J. C. López-Ardao, M. Fernández-Veiga, and C. López-García, "On the flexibility of M/G/∞ processes for modeling traffic correlations," in *International Teletraffic Congress*, 2007, vol. 4516 LNCS, pp. 284–294.

[21] R. D. Smith, "The dynamics of internet traffic: self-similarity, self-organization, and complex phenomena," *Adv. Complex Syst.*, vol. 14, no. 6, pp. 905–949, 2011.

[22] H. Sheng, Y. Q. Chen, and T. Qiu, "On the robustness of Hurst estimators," *IET Signal Process.*, vol. 5, no. 2, pp. 209–225, 2011.

[23] M. Pagano, "Self-similarity and long range dependence in teletraffic," in *International Asian School-Seminar Optimization Problems of Complex Systems*, 2019, pp. 125–130.

[24] C. Granero-Belinchón, S. G. Roux, and N. B. Garnier, "Information theory for non-stationary processes with stationary increments," *Entropy*, vol. 21, no. 12, p. 1223, 2019.

[25] D. B. Percival and A. T. Walden, "Spectral analysis for univariate time series," in *Cambridge Series in Statistical and Probabilistic Mathematics*, Cambridge, MA: Cambridge University Press, 2020.

[26] T. Karagiannis, M. Faloutsos, and M. Molle, "A user-friendly self-similarity analysis tool," *Comput. Commun. Rev.*, vol. 33, no. 3, pp. 81–93, 2003.

[27] T. Karagiannis, "The SELFIS tool," *Projects*. [Online]. Available: http://alumni.cs.ucr.edu/~tkarag/Selfis/Selfis.html.

[28] "Wireshark foundation," *Wireshark*. [Online]. Available: https://www.wireshark.org/.